\documentclass[trackchanges, twocolumn]{aastex7}

\shorttitle{ASHES XIII. CMF}
\shortauthors{K. Morii et al.}
\usepackage{textcomp}
\usepackage{comment}
\usepackage{rotating}
\usepackage{subcaption}
\usepackage{amsmath}

\newcommand{\x}{\mbox{$\times$}}

\begin{document}

\title{The ALMA Survey of 70 \textmu m Dark High-mass Clumps in Early Stages (ASHES).\\ XIII. Core Mass Function, Lifetime, and Growth of Prestellar Cores}

\author[orcid=0000-0002-6752-6061,sname='Morii']{Kaho Morii}
\altaffiliation{CfA Postdoctoral Fellow}
\affiliation{Center for Astrophysics $|$ Harvard \& Smithsonian, 60 Garden Street, Cambridge, MA 02138, USA}
\affiliation{Department of Astronomy, Graduate School of Science, The University of Tokyo, 7-3-1 Hongo, Bunkyo-ku, Tokyo 113-0033, Japan}
\affiliation{National Astronomical Observatory of Japan, National Institutes of Natural Sciences, 2-21-1 Osawa, Mitaka, Tokyo 181-8588, Japan}
\email[show]{kaho.morii@cfa.harvard.edu}  

\author[orcid=0000-0002-7125-7685]{Patricio Sanhueza}
\affiliation{Department of Astronomy, School of Science, The University of Tokyo, 7-3-1 Hongo, Bunkyo-ku, Tokyo 113-0033, Japan}
\email{psanhueza@astron.s.u-tokyo.ac.jp} 

\author[orcid=0000-0002-7125-7685]{Qizhou Zhang}
\affiliation{Center for Astrophysics $|$ Harvard \& Smithsonian, 60 Garden Street, Cambridge, MA 02138, USA}
\email{qzhang@cfa.harvard.edu} 

\author[0000-0002-6428-9806]{Giovanni Sabatini}
\affiliation{INAF, Osservatorio Astrofisico di Arcetri, Largo E. Fermi 5, I-50125, Firenze, Italy;}
\email{giovanni.sabatini@inaf.it}

\author[0000-0003-1275-5251]{Shanghuo Li}
\affil{School of Astronomy and Space Science, Nanjing University, 163 Xianlin Avenue, Nanjing, Jiangsu, 210023, China}
\affil{Key Laboratory of Modern Astronomy and Astrophysics, Nanjing University, Ministry of Education, Nanjing, Jiangsu, 210023, China}
\email{shli@nju.edu.cn} 

\author[0000-0003-3814-4424]{Fabien Louvet}
\email{fabien.louvet@univ-grenoble-alpes.fr}
\affiliation{Univ. Grenoble Alpes, CNRS, IPAG, 38000 Grenoble, France}

\author[orcid=0000-0002-1700-090X]{Henrik Beuther}
\affiliation{Max Planck Institute for Astronomy, K\"onigstuhl 17, 69117 Heidelberg, Germany}
\email{beuther@mpia.de}

\author[0000-0002-8250-6827]{Fernando A. Olguin}
\email{f.olguin@yukawa.kyoto-u.ac.jp}
\affiliation{Center for Gravitational Physics, Yukawa Institute for Theoretical Physics, Kyoto University, Kitashirakawa Oiwakecho, Sakyo-ku, Kyoto 606-8502, Japan}
\affiliation{National Astronomical Observatory of Japan, National Institutes of Natural Sciences, 2-21-1 Osawa, Mitaka, Tokyo 181-8588, Japan} 

\author[0000-0001-5461-1905]{Shuting Lin}
\affiliation{Department of Astronomy, Xiamen University, Zengcuo'an West Road, Xiamen, 361005, China}
\email{linst@stu.xmu.edu.cn}

\author[orcid=0000-0002-2149-2660]{Daniel Tafoya}
\affiliation{Department of Space, Earth, and Enviroment, Chalmers University of Technology, Onsala Space Observatory, 439 92 Onsala, Sweden}
\email{daniel.tafoya@chalmers.se}

\author[0000-0003-4521-7492]{Takeshi Sakai}
\affil{Graduate School of Informatics and Engineering, The University of Electro-Communications, Chofu, Tokyo 182-8585, Japan}
\email{takeshi.sakai@uec.ac.jp}

\author[0000-0003-2619-9305]{Xing Lu}
\affiliation{Shanghai Astronomical Observatory, Chinese Academy of Sciences, 80 Nandan Road, Shanghai 200030, P.\ R.\ China}
\affiliation{100101 Key Laboratory of Radio Astronomy and Technology (Chinese Academy of Sciences), A20 Datun Road, Chaoyang District, Beijing, 100101, P.\ R.\ China}
\email{xinglv.nju@gmail.com}

\author[orcid=0000-0001-5431-2294]{Fumitaka Nakamura}
\affiliation{National Astronomical Observatory of Japan, National Institutes of Natural Sciences, 2-21-1 Osawa, Mitaka, Tokyo 181-8588, Japan}
\affiliation{Department of Astronomy, Graduate School of Science, The University of Tokyo, 7-3-1 Hongo, Bunkyo-ku, Tokyo 113-0033, Japan}
\email{fumitaka.nakamura@nao.ac.jp} 

\begin{abstract}
The core mass function (CMF) of prestellar cores is essential for understanding the initial conditions of star and cluster formation. However, the universality of the CMF and its relationship to the initial mass function (IMF) remain unclear. 
We study the CMF in the earliest stage of high-mass star formation using 461 prestellar core candidates and 254 protostellar cores as a part of the ALMA Survey of 70 \textmu m Dark High-mass Clumps in Early Stages (ASHES). 
We find that prestellar core candidates tend to have lower masses than protostellar cores. 
We also find that the lifetime of prestellar cores is several times longer than the freefall time, although it approaches the freefall time as the core mass increases. 
The CMF, including both protostellar and prestellar cores, has a power-law slope of -2.05$\pm$0.04, shallower than Salpeter's IMF slope of -2.35. 
Conversely, the CMF of gravitationally bound, prestellar cores has a steeper slope (-2.32$\pm$0.30), indistinguishable from Salpeter's slope. 
This finding is consistent with observations in both low-mass star-forming regions and high-mass protoclusters, implying a universal core formation mechanism. 
The protostellar CMF with a larger maximum core mass can be reproduced by the prestellar CMF when an external gas infall is considered. 
The inferred mass infall rate is higher than the Bondi-Hoyle-Lyttleton accretion rate and follows a shallower mass dependence (smaller power-law index), more consistent with the tidal-lobe accretion. 
This may contribute to the evolution of CMFs seen in later stages. 
\end{abstract}


\section{Introduction} 
\label{sec:intro} 
The initial mass function of stars (IMF) describes the distribution of stellar masses at the time of their formation. Since the properties and evolution of stars are primarily influenced by their mass, the IMF is essential for understanding the structure of our universe. 
The universality of the IMF has been proposed from observations in the solar neighborhood \citep[][and reviewed by \citealt{Lee20}]{Salpeter55, Kroupa01, Chabrier03}. 
Specifically, these IMFs exhibit a power-law tail in the range of 0.4--10 $M_\odot$ characterized by a power index of $\alpha = -2.35$ \citep[$dN/dM \propto M^\alpha$, where $N$ is the number of cores in the mass bin and $M$ is the core mass;][]{Salpeter55}. 
Additionally, they show a log-normal distribution on the lower-mass end, with a peak around 0.1--0.2 $M_\odot$ \citep{Kroupa01, Chabrier03}. 
This universality across various star forming regions suggests that IMFs are determined by universal physical properties. 
Recent advancements using highly precise parallaxes and proper motions from the Gaia mission have allowed for more accurate measurement of the Milky Way's field IMF based on a much larger sample size \citep[e.g.,][]{Mor19-imf, Sollima19-imf}. 
These findings do not contradict the notion of a universal IMF. However, it has been suggested that the IMF may vary with environmental factors such as those found in low-metalicity environments or in the galactic center \citep[][]{Marks12, LeeHennebelle18}. In fact, deviation from a universal IMF, such as a top-heavy IMF, have  recently been observationally reported \citep{Hosek19, Li23-imf, KlessenGlover23-lowmetal}.  
This raises critical questions about what makes the IMF universal and how it varies under different environmental conditions. 

The core mass function (CMF) describes the mass distribution of cores, which are the progenitors of stars.  In particular, the mass function of prestellar cores has been expected as the potential progenitor of the IMF. 
The CMF observed in nearby low-mass star-forming regions resembles the IMF, shifted to higher masses when compared to the IMF, but maintaining a similar shape \citep[e.g.,][see more in a review by \citealt{Ballesteros-Paredes20}]{Andre10, Konyves15, Fiorellino21}. 
This suggests a constant core-to-star formation efficiency of $\sim$30--40\%. 

However, deriving the CMF from observations involves several challenges, including defining cores's boundary, achieving adequate spatial resolution, and estimating the core mass. 
Accurate core identification requires sufficient spatial resolution and high sensitivity, as highlighted in \citet{Louvet21}, especially to determine the CMF peak. 
Recently, \citet{Pelkonen21} suggest that the shapes of CMFs and IMFs can still resemble each other, even if there is no one-to-one correspondence between cores and stars, and that a significant fraction of the stellar mass originates from outside the initially bound cores.

With these considerations in mind, several studies have raised questions about the universality of CMFs based on observations in active cluster-forming regions, such as W43-MM1 \citep{Motte18_natas}, showing a shallower slope at the high mass end. 
More recently, the ALMA-IMF Large Program \citep[e.g.,][]{Motte22-ALMAIMF,Ginsburg22-ALMAIMF} conducted high-resolution ALMA observations toward 15 massive protoclusters and increased the sample of top-heavy CMFs with a consistent methodology to derive CMFs \citep{Louvet24}. 
\citet{Pouteau23}  investigated the evolution of CMFs in the W43-MM2 and MM3 clusters, suggesting a transition from a Salpeter-like slope to a top-heavy slope as star formation becomes more active. 
Furthermore, \citet{Nony23} categorized cores as prestellar or protostellar based on the presence of molecular outflows, finding that the prestellar CMF follows a Salpeter-like distribution while the protostellar CMF exhibits a shallower slope. 
These studies highlight how radio interferometers enable to explore the CMF not only in our solar neighborhood but also in more distant, massive protoclusters. 

Moreover, recent high-resolution observations have also revealed the inner structures of cores, such as fragments, asymmetric structures, and disk-like structures \citep[e.g.,][; Luo et al., Submitted]{Beuther21, Sanhueza21, Sanhueza25,Olguin22,Olguin23,Li24_natureas, Sanna25, Olguin25-arxiv, Mai25-arxiv}. 
This indicates a complex mass delivery process from cores to stars, especially in cluster-forming regions. A constant core-to-star formation efficiency and a simple shift in masses from the CMF to the IMF are likely oversimplifications based on the recent high-angular resolution observations. 
Nevertheless, the CMF, particularly the prestellar CMF, is an important key property characterizing the initial conditions of stellar cluster formation. 

Infrared dark clouds (IRDCs) are dark clouds that appear in silhouette in infrared images due to the background galactic plane emission and the presence of dense gas. 
These clouds are believed to be the birthplace of high-mass stars or stellar clusters because of the lack of bright infrared emission \citep{Sridharan05, Rathborne06, BerginTafalla07,Chambers09,Sanhueza12, Sabatini19}. 
IRDCs tend to be identified before widespread star formation is evident, and once the high-mass stars turn on, they disrupt the surrounding environment through intense radiation, stellar winds, and the formation of HII regions. 
Therefore, studying IRDCs is crucial for understanding the evolution of high-mass star formation \citep[e.g.,][]{Csengeri17b_progenitor,Sanhueza13,Sanhueza17}. 
The CMF within IRDCs likely reflects the mass distribution of cores during the early stages of cluster evolution. 

However, IRDCs are typically located far from the Sun (i.e., $>$a few kpc), and the CMFs that have been produced so far often suffer from small sample sizes or insufficient spatial resolution to accurately identify cores \citep{Ohashi16, Kong19_core, Liu18}.  
Additionally, the complexity introduced by molecular outflows in these cluster-forming regions complicates the determination of the mass function of prestellar cores. 
In the ALMA Survey of 70 \textmu m Dark High-Mass Clumps in Early Stages (ASHES), we obtained the large, extremely well-characterized population of cores embedded in the IRDCs \citep[e.g.,][]{Sanhueza19, Li20, Tafoya21, Morii21, Li22, Sabatini22, Sakai22, Li23, Morii23, Izumi24, Morii24, Lin25}. 
We identified 839 cores and carefully classified their evolutionary stages by detecting molecular outflows and warm gas tracers \citep{Sanhueza19, Li20, Morii21, Morii23, Izumi24}, as well as assessing their gravitational stability through virial analysis \citep{Li23, Morii24}. 

In this paper, we present the mass function of cores identified in the ASHES, allowing for comparison with those from other star-forming regions. 
This paper is constructed as follows: Section~\ref{sec:obs} describes the observation, data properties, and core sample. Section~\ref{sec:results} explains how to derive the CMF, and display the produced CMF. 
We discuss the slope of the CMF and the lifetime of prestellar cores in Section~\ref{sec:discussion}. Our conclusion is listed in Section~\ref{sec:conclusion}. 

\section{Observations and Data} \label{sec:obs} 
We have used the observational data from the ASHES project. 
Observations were carried out with ALMA in Band 6 ($\sim$224 GHz;  $\sim$1.3 mm) through three cycles: Cycle 3 (2015.1.01539.S, PI: P. Sanhueza), Cycle 5 (2017.1.00716.S, PI: P. Sanhueza), and Cycle 6 (2018.1.00192.S, PI: P. Sanhueza). 
The ASHES-sample consists of thirty-nine 70\,\textmu m-dark IRDC clumps with the potential for high-mass star formation \citep[see Table 1 in][]{Morii23}. 
The data were taken with the main 12-m array and the Atacama Compact Array (ACA), including both the 7-m array and Total Power (TP; for spectral line data only). 
The average 1$\sigma$ rms noise level of the continuum images was $\sim$0.094 mJy beam$^{-1}$ with a beam size of $\sim$1\farcs2. 
\citet{Sanhueza19} and \citet{Morii23} describe the detailed ASHES program and the observation setup.  
In this paper, we use CO ($J$=2--1), H$_2$CO ($J_{K_a, K_c} = 3_{2,2}-2_{2,1}, 3_{2,1}-2_{2,0}$), CH$_3$OH ($J_K = 4_2-3_1$), and HC$_3$N ($J= 24-23$) for identifying protostellar cores. 
Their velocity resolutions are $\sim$1.3 km\,s$^{-1}$. DCO$^+$ ($J=3-2$) and N$_2$D$^+$ ($J=3-2$) with velocity resolutions of 0.17 km\,s$^{-1}$ are used for the virial analysis. 
Those lines are produced by combining the 12 m array data with the 7 m array data. We also use TP data of the C$^{18}$O ($J= 2-1$) to estimate the velocity dispersion of the clumps, which has a velocity resolution of 1.3 km\,s$^{-1}$ and an angular resolution of 29\farcs6.

By applying the dendrogram technique, implemented in \texttt{astrodendro} \citep{Rosolowsky08}, to the 1.3 mm continuum images of 39 regions, we have identified 839 cores \citep{Morii23}. 
We classified cores into prestellar core candidates and protostellar cores using the molecular lines. 
Detailed information and line setups are described in \citet{Morii24}. 
Among 839 cores, 103 cores (12.3\%) are associated with molecular outflows detected in CO ($J$=2--1), and 222 cores (26.4\%) are not associated with CO outflows but have detections of at least one warm gas tracer, such as H$_2$CO ($J_{K_a, K_c} = 3_{2,2}-2_{2,1}, 3_{2,1}-2_{2,0}$), CH$_3$OH ($J_K = 4_2-3_1$), or HC$_3$N ($J= 24-23$), with $E_{\rm u}/k$ of 68.09, 68.11, 45.46, and 130.98 K, respectively. 
These warm gas tracers have similar spatial distributions as outflows and the derived rotational temperature is high ($\gtrsim$50 K), implying ongoing star formation activity \citep{Morii21, Izumi24}. 
Hereafter, they are named as outflow cores and warm cores, respectively, or protostellar cores by combining both outflow and warm cores.   
The remaining 514 cores (61.3\%) without any detections of star formation signatures are prestellar core candidates.  
We also calculated the virial parameters with the velocity dispersion estimated from the dense gas tracers of DCO$^+$  or N$_2$D$^+$ $J=3-2$ since their similar spatial distributions to the dust continuum emission and their high critical density indicates that they trace the dense cores \citep{Sakai22, Li23, Morii21}. 
Among 839 cores, either of the lines is detected from 492 cores (58.6\%), and 340 cores (69.1\%) have virial parameter smaller than 2, classified as gravitationally bound cores ignoring external pressure and the magnetic field \citep{Morii24}. 

\section{Results} \label{sec:results}
\subsection{Core Mass} \label{sec:mass}
The core masses are derived from the 1.3 mm dust continuum emission adopting the optically thin assumption in
\cite{Morii23} as 
\begin{equation}
    M_\mathrm{core} = \mathbb{R}\frac{d^2 F_\nu}{\kappa_\nu B_\nu (T_\mathrm{dust})},
\label{equ:Mass}
\end{equation}
where $\mathbb{R}$ is the gas-to-dust mass ratio, $F_\nu$ is flux density, $\kappa_\nu$ is the dust absorption coefficient, $d$ is the distance to the source, and $B_\nu$ is the Planck function for a dust temperature $T_\mathrm{dust}$. 
For cores in eleven ASHES clumps, we used the rotational temperatures derived from NH$_3$ observations at $\sim$5\arcsec \citep[][]{Li23, Allingham24}.  
For the remaining twenty-eight clumps, we adopted the clump temperature for the core mass calculation \citep{Guzman15}. 
Column 3 in Table~4 in \citet{Morii23} shows the temperature used for each core in the mass estimation. 
We adopt the standard gas-to-dust mass ratio of 100, and a dust opacity of 0.9\,cm$^2$\,g$^{-1}$ \citep{Ossenkopf94}. 

We checked the optical depth at 1.3 mm ($\tau_{\rm 1.3mm}$) at first order by taking the ratio of $I_{\rm 1.3mm, peak} / \Omega$ to $B_{\rm 1.3mm}(T_{\rm dust})$, corresponding to 1-exp(-$\tau_{\rm 1.3mm}$), where $I_{\rm 1.3mm, peak}$ is the measured peak intensity, $\Omega$ is the beam solid angle, as introduced in Appendix B of \citet{Pouteau22}. 
The cores have ratios smaller than 0.08 (median of 0.001), which is in agreement with the optically thin condition.    

The derived core mass ranges from 0.06 to 81 $M_\odot$. 
The uncertainties of mass is $\sim$50\%, largely contributed from the gas-to-dust mass ratio and the dust opacity \citep[see][for more information]{Morii23}.

\begin{figure*}
\centering
    \includegraphics[width=0.99\linewidth]{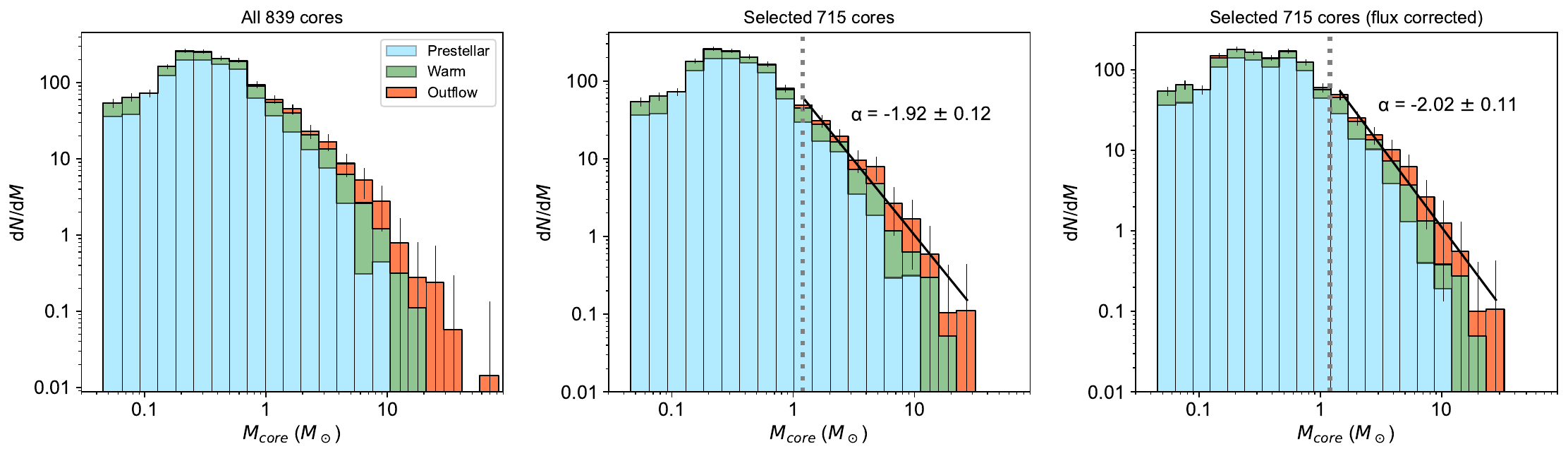}
    \caption{CMF of (left) the full sample, (middle) the selected sample, and (right) the selected sample corrected by the recoverable flux. The three different colors represent the evolutionary stages of cores: prestellar core candidates (light blue), warm core (green), and outflow core (orange). 
    Each bin in the histogram represents the combined frequency from several stages, stacked to show the relative contribution of each dataset to the total count per bin. The worst 90\% completeness (1.2 $M_\odot$) is plotted in the middle and right panels. The results of power-law fit above 1.2 $M_\odot$ are also overplotted. }
    \label{fig:cmf-all}
\end{figure*}

The CMF with a differential histogram of all 839 cores is shown in the left panel of Figure~\ref{fig:cmf-all}.
The bin size was determined with a task in \texttt{numpy} of \textit{histogram bin edges} as Freedman Diaconis Estimator (fd), which takes into account data size.  
The colors light blue, green, and orange represent prestellar core candidates, warm cores, and outflow cores, respectively. 
Here, each bin represents the combined frequency from three evolutionary stages, stacked to show the relative contribution to the total count per bin. 
This plot clearly shows the increase of the percentage of protostellar cores as core mass increases. 
Although the minimum mass of outflow and warm cores is small, 0.2 $M_\odot$ and 0.06 $M_\odot$, respectively, the mass of protostellar cores are generally larger than those of prestellar core candidates. 

\subsection{Completeness} \label{sec:completeness}
\begin{figure}[htbp]
    \centering
    \begin{subfigure}{\linewidth}
    \leavevmode
        \centering
        \includegraphics[width=\linewidth]{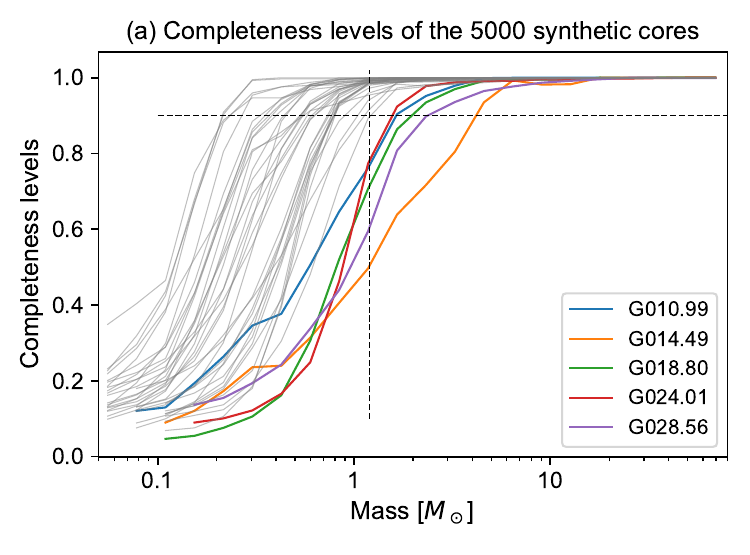}
    \end{subfigure}
    \vfill
    \begin{subfigure}{\linewidth}
    \leavevmode
        \centering
        \includegraphics[width=\linewidth]{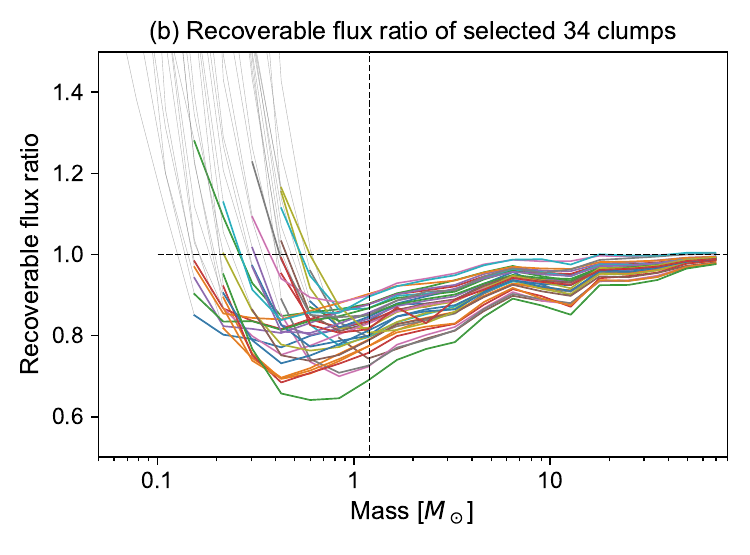}
    \end{subfigure}
    \caption{Completeness analysis of synthetic cores embedded in the continuum image. Regions with completeness levels below 90\% at 1.2 $M_\odot$ are highlighted as colors in (a). In (b), the clumps with completeness levels below 90\% at 1.2 $M_\odot$ are excluded. The lines are colored in for the mass range with the completeness level higher than 50\%, which are used for flux correction. } 
    \label{fig:CMF_completeness}
\end{figure}

We calculated the mass completeness by inserting an artificial Gaussian core into the continuum images and varying the core mass. 
For each mass bin, the flux of the Gaussian core was calculated from Equation~\ref{equ:Mass} considering the primary beam correction.  
The inserted position of core center was uniformly random in a region where the primary beam response was higher than 40\% to avoid cores located at the edge, consistent with the core identification process.  
After inserting a core, we applied the dendrogram algorithm to the image and checked if the inserted core was identified as a $\mathtt{leaf}$ whose peak position is consistent with the inserted position within the beam size. 
We repeated the test 5000 times and calculated the mean detection rate per mass bin from region to region to determine the 90\% completeness level.  
Figure~\ref{fig:CMF_completeness} (a) shows the fraction of detected cores as a function of core mass. 
The completeness level ranges from 0.3--4.5 $M_\odot$ with a median of $\sim$0.6 $M_\odot$. 
More deficient completeness comes from the worse mass sensitivities.    
To reduce these effects among the sample, we excluded five clumps with the lowest completeness (see the colored lines in the figure)\footnote{The five clumps with low completeness levels are G010.991–00.082, G014.492–00.139, G018.801--00.297, G024.010+00.489, and G028.564--00.236.}. 
After this selection, our sample ensures a 90\% completeness level above 1.2 $M_\odot$, resulting in a final sample of 715 cores for the CMF analysis. 

We also measured how much flux was recovered by the dendrogram algorithm  by comparing the flux of leaves identified by \texttt{astrodendro} and the inserted flux. 
The recovered flux profile is shown in Figure~\ref{fig:CMF_completeness} (b), showing that above 1.2 $M_\odot$, the recovered flux ranges from $\sim$0.7 to $\sim$1, but most closely resembles unity. 
The ratio of the identified and inserted flux can be higher than 1 if a core is inserted on the existing emission such as filament or another source, which may happens in the heavily clustered regions. 
Using this flux recovery profile, we corrected the observed masses by dividing them by the corresponding recovery fraction in each mass bin and region. 
The correction has been applied to the mass range of each core where the completeness level is higher than 0.5, colored in Figure~\ref{fig:CMF_completeness}. 
This correction slightly increases the masses of low-mass cores. 

The CMFs of the selected 715 cores before and after flux correction are shown in the middle and right panels of Figure~\ref{fig:cmf-all}.  
Among the 715 cores in the middle panel, the numbers of prestellar core candidates, warm cores, and outflow cores are 461, 178, and 76, respectively. 
The worst 90\% completeness level (1.2 $M_\odot$) is plotted as dotted lines. 
The middle panel of Figure~\ref{fig:cmf-all} indicates that the completeness limit is above the detection limit of outflows and warm lines (see Appendix~\ref{sec:Appendix_Ncore}) and also the protostellar core fraction increases as the core mass. 
A comparison between the left and middle panels reveals that the excluded sample contains some of the most massive cores\footnote{This is mostly due to G028.564--00.236, which host 5 out of 7 cores with masses of $>$25$M_\odot$}, but the distributions of the intermediate-mass range, which dominate the fitting result, are similar. 

We applied the power-law fitting above 1.2 $M_\odot$ in the selected sample case. 
As expected from the recovered flux plot, the flux correction mostly affects low-mass cores near 1 $M_\odot$, and it makes a slightly steeper slope from -1.92$\pm$0.12 to -2.02$\pm$0.11.  
Here, the uncertainties includes the mass uncertainties and the fitting uncertainties. 
Even considering the uncertainties, these slopes are slightly shallower than -2.35, consistent with the first attempts of studying the CMF in the ASHES pilot survey \citep{Sanhueza19} and high-mass protoclusters \citep[e.g.,][]{Motte18_natas, Kong19_core, Moser20, Louvet24}. 
To confirm our results are independent of the core identification method adopted (astrodendro), we ran \texttt{getsf} \citep{Menshchikov13, Menshchikov21}, as described in Appendix~\ref{sec:Appendix_coremethod}, and obtain a consistent result. 

\subsection{Prestellar CMF} 
\begin{figure*} 
    \centering
    \includegraphics[width=0.99\linewidth]{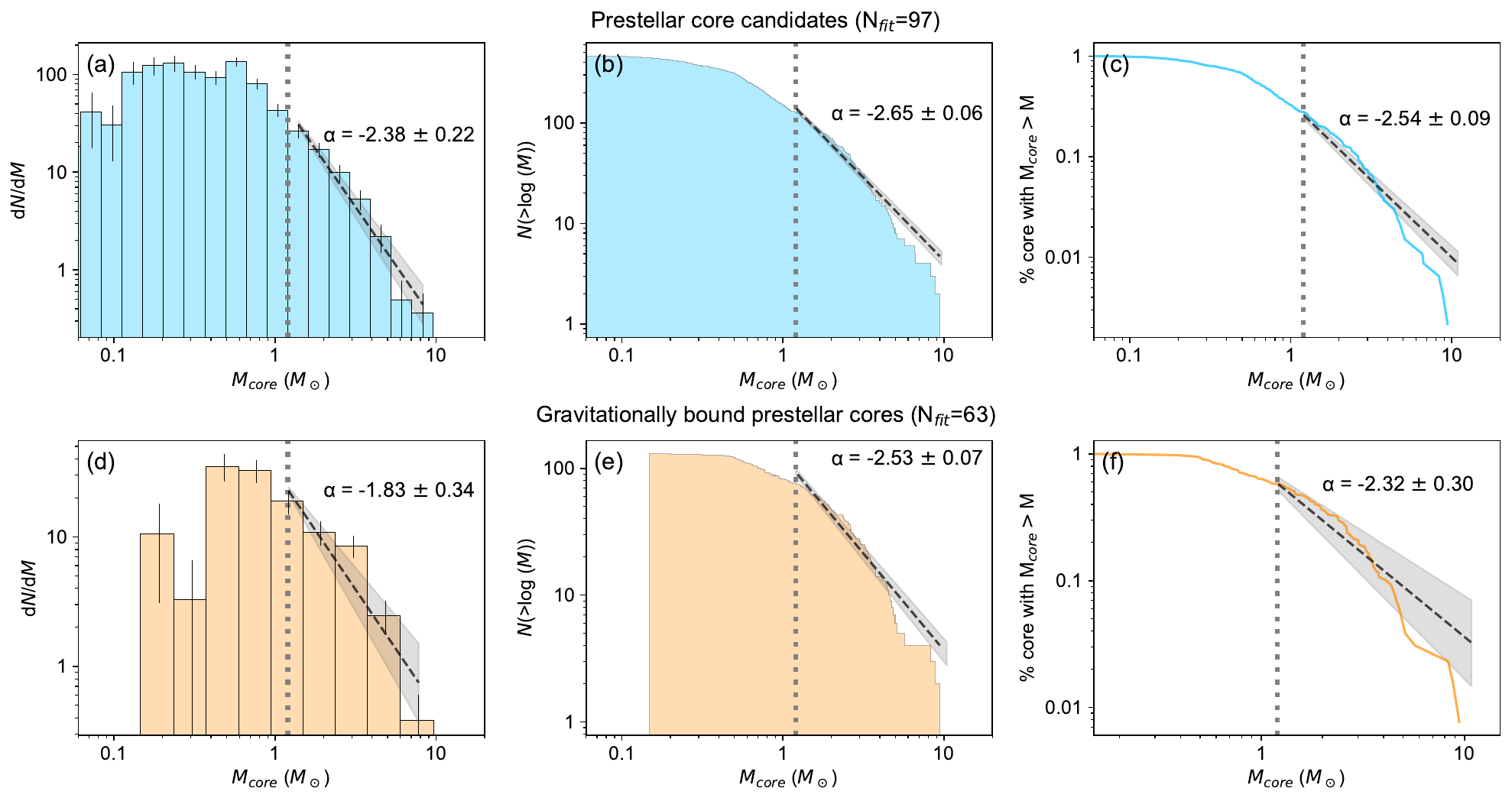}
    \caption{Core mass distributions of (a-c) prestellar core candidates and (d-f) gravitationally bound prestellar cores. Three columns show different plots in the form of (a, d)  differential CMFs, (b,e) cumulative CMFs, and (c, f) cumulative percentage of cores with $M_{\rm core}>M$, respectively. The best-fitting power-law index is plotted in each panel. The number of cores above 1.2 $M_\odot$ in each sample is shown. }
    \label{fig:cmf_prebound}
\end{figure*}

Prestellar cores are cores on the verge of gravitational collapse, prior to the formation of protostars. 
Therefore, the prestellar mass function likely traces the initial conditions of star formation. 
By comparing its properties in various environments such as low-mass and high-mass star-forming regions, outer galaxy clouds, galactic center regions, and also with the IMF, we can investigate the universality of star formation in the galaxy. 
Our prestellar core sample consists of 461 cores without any signs of protostars detected by ALMA such as molecular outflows and warm gas tracers.  
It includes 130 gravitationally bound cores based on virial analysis.  
The CMF, in differential form, of both all prestellar core candidates and gravitationally bound prestellar cores is shown in Figures~\ref{fig:cmf_prebound} (a) and (d). 

In addition to the standard differential histogram, to avoid arbitrariness of binning and to directly compare with other references, we also derived the complementary cumulative distribution with variable-size bins, Figure~\ref{fig:cmf_prebound} (b) and 3(e), following the recommendations of \citet[][]{MaizApellaniz05} and \citet{Reid06}, the same method as \citet{Motte18_natas}. 
Also, we performed the fitting using maximum likelihood estimates (MLE; Figures~\ref{fig:cmf_prebound} (c) and (f)) as developed by \citet{Clauset09}, following one of the ALMA-IMF studies by \citet{Louvet24} and implemented in the Python package, \texttt{powerlaw} \citep{Alstott14}. 
MLE determines the best-fit power-law slope directly from the unbinned data by maximizing the likelihood of the observed distribution, avoiding biases from histogram binning. 
Figure~\ref{fig:cmf_prebound} shows all the CMFs of the prestellar core candidates (top) and gravitationally bound prestellar cores (bottom). 

The black lines represent the best fit of the power law index above the lowest 90\% completeness level of 1.2 $M_\odot$, $dN/dM \propto M^\alpha$, $N(> \mathrm{log} M) \propto M^{\alpha +1}$, or $P(m) = - \frac{\alpha +1}{M_{\rm min}} \left(\frac{m}{M_{\rm min}} \right)^{\alpha}$.  
The gray-shaded regions show 1$\sigma$ uncertainty on alpha, which comes from the fitting uncertainties and the mass uncertainties. 
The latter is estimated from the fitting of the simulated CMFs.  
Considering a 50\% uncertainty in core masses (Section~\ref{sec:mass}), we performed a Monte Carlo simulation to assess how the CMF changes. 
We multiplied a random value in the range of 0.5--1.5 for each core mass in one trial and produced the CMF. We repeated this simulation 1000 times and calculated a 1 sigma value coming from the mass uncertainty.  

All distributions show power-law tails in the mass range of at least 1--5 $M_\odot$, then deviate toward a steeper decline at higher masses.   
This figure shows that the power-law index $\alpha$ slightly depends on the representations of the mass distribution in such a limited sample size at the high-mass end. 
The best fit $\alpha$ was -2.65 to -2.38 for prestellar core candidates (including bound and unbound cores) and -2.53 to -1.83 for bound prestellar cores. 
When comparing the CMFs derived from the same method, we find that the bound core-limited CMFs have a slightly shallower slope, but they still remain around -2.35. 

\section{Discussion} \label{sec:discussion}
\subsection{Power-law Index of CMFs}
\begin{figure}
    \centering
    \includegraphics[width=\linewidth]{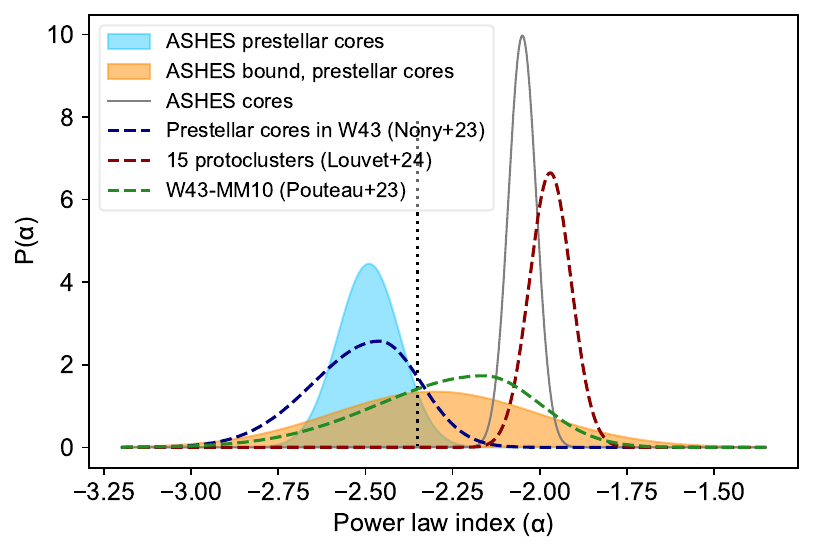}
    \caption{The power-law index ($\alpha$ of $dN/dM \propto M^\alpha$) best-fitted by the CMFs in the form of a cumulative percentage of cores with $M_{\rm core}>M$. 
    Our results of the prestellar-CMF and prestellar-bound CMF are shown as light blue and orange colors, respectively. 
    For comparison, the index from the CMF of all-ASHES cores (gray), prestellar cores in W43 \citep[blue;][]{Nony23}, 15 clusters in ALMA-IMF \citep[dark red;][]{Louvet24}, and young cluster W43-MM10 \citep[green;][]{Pouteau23} are included. The vertical black dotted line represents -2.35. }
    \label{fig:cmf-slope}
\end{figure}

Figure~\ref{fig:cmf-slope} summarizes the best-fit power-law index of CMF of prestellar core candidates (light blue) and bound-prestellar CMF (orange). 
The shaded probability density function is built from the fitting of the cumulative percentage distributions using the MLE method, corresponding to Figure~\ref{fig:cmf_prebound} (c) and (f).  
The value of $\alpha=-2.35$ is indicated by the vertical line. 
The light blue and orange distributions are both located near or at values smaller than -2.35. 

For comparison, we plotted the index and the uncertainties estimated in some studies of the ALMA-IMF project \citep{Nony23, Pouteau23, Louvet24}. 
It should be noted that the completeness and the fitting range (i.e., the minimum mass used for fitting) is comparable, and all values in this plot are derived using the same fitting methods. 
One of the published ALMA-IMF papers, \citet{Nony23}, only present the CMF of prestellar cores. 
They show that the slope of prestellar CMF is steeper than that of protostellar CMF, and it was $\alpha=-2.46 ^{+0.12}_{-0.19}$ (blue dashed line). 
Upon a direct comparison of the blue dashed line with our distribution, the best-fit index is well consistent, slightly below -2.35. 

In addition to the prestellar core mass function, we also compare the CMF that includes both prestellar and protostellar cores, shown in green, red, and gray curves. 
\citet{Pouteau23} produced CMFs in subregions of the W43 protocluster to study their evolution. 
Here, we plot one of the measured values, $\alpha=-2.16^{+0.16}_{-0.30}$ from W43-MM10 (green dashed line), which is likely in an early evolutionary stage.
\citet{Louvet24} derived CMFs by combining all the cores identified in 15 protoclusters and estimated $\alpha=-1.97 \pm 0.06$ (red dashed line). 
Since they used the MLE method, to directly compare with our sample, we also plot the best-fit index derived in the same way as a gray line ($\alpha=$-2.05$\pm$0.04). 
The gray line takes its peak similar to the green but smaller than the red line. 
Our sample consists of cores embedded in 70 \textmu m-dark clumps that are likely in a very early evolutionary stage. W43-MM10 is also in similar, relatively quiescent region. 
This trend, from early evolutionary stages (quiescent, 70 \textmu m-dark regions like our sample and W43‑MM10) to more evolved, active cluster-forming stages (like those in \citealt{Louvet24}), is consistent with the ALMA‑IMF \citep{Pouteau23} results. Furthermore, it is also in agreement with the ALMAGAL survey \citep{Coletta25-almagal} which suggest an increase in the power-law index during cluster evolution.

In summary, Figure~\ref{fig:cmf-slope} reveals two key findings. 
First, the comparison of the CMF of the full sample (including prestellar and protostellar cores) between quiescent regions (gray and green) and protoclusters (red) tells that the slope becomes shallower as protoclusters evolve. 
This implies that the full CMF reflects star formation activity and the formation of high-mass cores (see Section~\ref{sec:core-growth} for further discussion). 
Second, the prestellar CMFs (light blue, orange, and blue) has a slope consistent with the Salpeter IMF, and this result has been reported across various evolutionary stages \citep[][and also \citealt{Suarez21, Cheng24}]{Nony23} and even in low-mass star-forming regions. 
This suggests that the prestellar core formation mechanism may be a universal process throughout the galaxy, largely independent of the surrounding star-forming activity. 

\subsection{Lifetime of Prestellar Cores} \label{subsec:lifetime} 
The ASHES core sample, the large, well-characterized sample of cores embedded in IRDCs, 
allows us to constrain the lifetime of prestellar cores in cluster-forming regions. 

Whether prestellar cores follow fast collapse or slow collapse models is one of the important questions in star formation. 
The former case expects $\lesssim$1 Myr, comparable to the freefall time of cores, where cores form through gravo-turbulent fragmentation \citep[e.g.,][]{Hennebelle12} and the latter expects a longer lifetime of 1--10 Myr, several times longer than the freefall time, due to the support from magnetic fields and turbulence \citep{Shu87}. 
In nearby star-forming regions, several attempts have been made to estimate the lifetime of prestellar cores such as careful physical and chemical modeling \citep{Lin20, Lin24} and also in a statistical way counting the number of the prestellar and protostellar cores  \citep[i.e., statistical lifetime][]{Beichman86, LeeMyers99, Jessop00, Ohnishi02, Konyves15, Tokuda20, Takemura23}. 
Assuming a constant star formation rate throughout the cloud's lifetime, the statistical lifetime can be derived as
\begin{equation}
    t_{\rm pre} = \frac{\rm Number\,of\,prestellar\,core}{\rm Number\,of\,protostellar\,core} \times t_{\rm proto}, 
    \label{equ:lifetime}
\end{equation}
where $t_{\rm proto}$ is the reference lifetime. 
In the low-mass regime, or core density range of 10$^4$--10$^5$ cm$^{-3}$, the prestellar core lifetime has been estimated to be $\sim$1.2 Myr, a few times the core freefall time, $t_{\rm ff}$, which implies nonthermal support slow down collapse \citep[e.g., turbulence and magnetic fields;][]{Ward-Thompson07, Konyves15, Tokuda20, Takemura23}. 

Here, we estimate the statitical lifetime of prestellar cores in the ASHES sample. 
Due to the lack of a formal definition of classes based on Spectral Energy Distributions (SEDs) in high-mass star-forming regions, and following studies that calculated the relative number of protostellar cores with HII regions \citep{Tige17, Motte18}, we adopt 0.2 Myr as the protostellar lifetime, $t_{\rm proto}$. 
As discussed in \citet{Valeille-Manet25}, this has an uncertainty of about a factor of two. 
By inserting the total number of bound, prestellar cores and the number of bound, protostellar cores (including both warm cores and outflow cores) into Equation~\ref{equ:lifetime}, we estimate the lifetime of prestellar cores to be 0.18 Myr.  
This value is $\sim$6 times longer than the freefall time of the ASHES cores, which is approximately 3$\times$10$^4$ yr.     

\begin{figure}
    \centering
    \includegraphics[width=\linewidth]{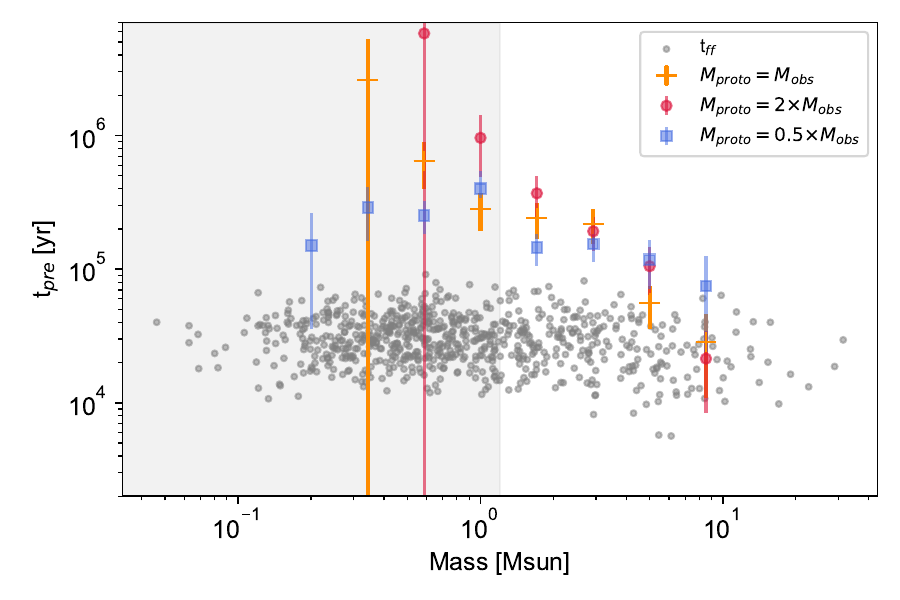}
    \caption{Prestellar lifetime estimated from the number ratio of bound prestellar and  bound protostellar cores against core mass. Orange crosses show the ratio using the observed core mass. The red and blue markers show the case assuming the accretion and external infall by multiply the protostellar core masses by 2 and 0.5, respectively. Each corresponds to accretion- or external infall-dominated case. The error bar represents the statistical error in each mass bin. Gray circles represent the freefall time of each cores. The completeness limit is higlighted as the shaded area. }
    \label{fig:lifetime}
\end{figure}

The large core sample also allows us to investigate a possible differences of the core lifetime with respect to the core mass. 
Figure~\ref{fig:lifetime} illustrates the estimated core lifetime as a function of mass, represented by orange crosses. 
The data points were obtained by first creating equally spaced mass bins on a logarithmic scale.
In each mass bin, we counted the number of prestellar and protostellar cores and applied Equation~\ref{equ:lifetime}. Here, we limited the sample only to gravitationally bound cores. 
The resulting plot clearly demonstrates the mass dependency of the core lifetime. 
Gray circles represent the freefall time of each core.
As the core mass increases, the lifetime decreases from $\sim$1 Myr to $\lesssim$0.1 Myr, approaching their corresponding freefall time. 

Recently, the ALMA-IMF project reported the lifetime of high-mass prestellar core candidates ($>$8 $M_\odot$), as $\sim$0.12--0.24 Myr or $\sim$10$\times t_{\rm ff}$ assuming a protostellar lifetime of 0.3 Myr \citep{Valeille-Manet25}. 
However, even though our sample at $\sim$10 $M_\odot$ is not large, the estimated lifetime is much shorter than 0.1 Myr. 
The non-detection of high-mass prestellar cores with masses larger than 10 $M_\odot$ and lower percentage of prestellar cores in the intermediate-mass range ($>5\,M_\odot$) in ASHES lead to a shorter lifetime. 
Even when comparing the number of prestellar core candidates and cores associated with outflows (i.e., excluding warm cores from the protostellar core sample) or including warm cores in the prestellar core sample, the derived lifetime is still less than 0.1 Myr at 10 $M_\odot$. 
In the lower mass range, the derived values are a few times longer than the freefall time, which is consistent with those in nearby regions, including Orion \citep{Konyves15, Takemura23}. 

Up to this point, our analysis of core lifetimes as a function of mass has implicitly assumed that core mass remains constant over time. This assumption is equivalent to a scenario where gas accretion onto the central protostar is perfectly balanced by replenishment from the surroundings. 
However, the evolution of core mass during star formation involves both accretion onto the central object (leading to a decrease in core mass) and gas infall from the surrounding clump (increasing core mass), and their rate can be different to each other.  
We now consider these two alternative scenarios to better understand the mass dependency of core lifetimes.

First, let us consider a scenario where the core mass decreases over time due to accretion onto the forming protostar. 
In this case, the initial core mass would have been larger than the currently observed core mass. 
For example, if half of the initial core mass has already been accreted onto the protostar, then their initial core mass would be twice the currently observed core mass ($M_{\rm proto}=2 \times M_{\rm obs}$). 
The red circles in Figure~\ref{fig:lifetime} represent the core lifetimes estimated under this assumption. 
Although this scenario suggests a more rapid mass decrease, it does not significantly alter our primary conclusion: more massive cores still exhibit lifetimes of a few times 10$^4$ yr. 
This is comparable to the lifetime of the 70 \textmu m dark phase \citep{Motte07, Tackenberg12, Csengeri14, Urquhart18, Sabatini21}, implying a rapid mode of star formation. 

Alternatively, the core mass may increase during its evolution as discussed in Section~\ref{sec:core-growth}. 
For instance, this considers the case in which the core continues to accrete material from the surrounding clump at a higher rate than the one from which the star accretes from the core. 
Indeed, \citet{Coletta25-almagal} revealed the increase of core masses as clump evolution as a part of the ALMAGAL project, implying such core growth \citep[see also][]{Xu24, Morii25}.  
Under this growth scenario, the initial core mass would be smaller than the currently observed core mass. 
For example, if the initial core mass was half of the observed mass ($M_{\rm proto}=0.5 \times M_{\rm obs}$), the estimated values are shown in Figure~\ref{fig:lifetime} as blue squares. 
The derived prestellar lifetime at $M>$5 $M_\odot$ becomes longer, and around 0.1 Myr. 
In this case, even in the high-mass range, the lifetime may be several times longer than the freefall time. 
This is similar to the ALMA-IMF estimation \citep{Valeille-Manet25}, although it is subject to considerable uncertainties. 

Overall, the statistically derived timescale of prestellar cores in the mass range of 1--5 $M_\odot$ is generally around 0.1 Myr. This is likely a few times longer than the cores' freefall time, even considering the accretion- or external-infall-dominated cases. 
It aligns with the statistical lifetime estimated in low-mass star-forming regions derived similarly. 
\citet{Bovino21} revealed a comparable prestellar lifetime using a chemical clock that combines the observed pD$_2$H$^+$/oH$_2$D$^+$ ratio with numerical calculations. 
Although their calculation assumes a low-mass star formation environment, the derived value of a few to 10 times the freefall time is consistent with ours. 
At $\sim$5--10 $M_\odot$, the prestellar lifetime may be close to cores' freefall time without a significant infall contribution. Further investigation using the deuterium ratio in IRDCs, as attempted by \citet{Redaelli21, Redaelli22}, would provide a different perspective to constrain the lifetime. 

One caveat of this analysis is that we assumed that the star formation rate (SFR) remains constant within each clump. 
If the SFR at the current stage is low, the derived lifetime may be overestimated. 
In such cases, the lifetime of a prestellar core is generally longer than its freefall time, even for cores with a mass of $\sim$10 $M_\odot$. 

\subsection{Core Growth and CMF} \label{sec:core-growth}
\begin{figure}
    \centering
    \includegraphics[width=0.97\linewidth]{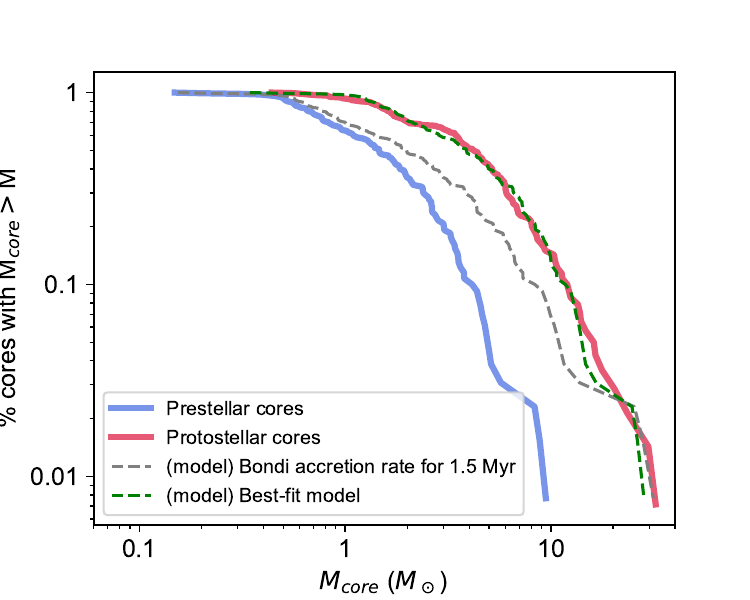}
    \caption{Accumulated percentage derived from gravitationally bound, prestellar cores (blue) and bound, protostellar cores (red), compared with the CMFs assuming external infall due to Bondi-Hoyle-Lyttleton accretion for 1.5 Myr (dashed gray), and the best-fit model (dashed green; see the main text). The blue line is the same with Figure~\ref{fig:cmf_prebound} (f). }
    \label{fig:cmf_evolution}
\end{figure}

The ASHES project has investigated core properties in the very early stages and revealed a lack of high-mass cores in most targets \citep{Morii23}, implying the need for core growth.
This conclusion is supported not only by ASHES but also by recent ALMA surveys of high-mass star-forming regions, such as QUARKS (Querying Underlying mechanisms of massive star formation with ALMA-Resolved gas Kinematics and Structures; \citealt{Xu24}) and SQUALO (QUiescent And Luminous Objects; \citealt{Traficante23}), as well as large programs like ALMA-IMF \citep{Motte22-ALMAIMF} and ALMAGAL \citep{Molinari25-almagalI, Coletta25-almagal}.
Given the dense, massive, and clustered environments where high-mass stars form, it is natural to expect that gas initially unbound to cores continues to infall from the surroundings, even after protostellar formation. 
As seen in Figure~\ref{fig:cmf-all}, protostellar cores tend to have larger masses than prestellar cores. 
Figure~\ref{fig:cmf_evolution} shows the CMF of gravitationally bound prestellar cores and that of bound protostellar cores as blue and red lines, respectively. 
Again, it shows that mass function of protostellar cores
is shallower with a larger maximum core mass.   
Similar features have also been reported in more massive star cluster-forming regions \citep{Nony23}. 

Here, we investigate the effect of external infall or core growth on the CMF, as well as the amount of external infall needed to reproduce the observed protostellar CMF from the prestellar CMF.
First, we estimate the Bondi-Hoyle-Lyttleton (BHL) accretion as 
\begin{eqnarray}
    \nonumber
    \dot{M}_{\rm Bondi} &=& \frac{\pi \rho G^2 M^2}{\sigma^3} \\
    \nonumber
    &\sim& 1.65\times 10^{-7} \left( \frac{n_{\rm cl}}{4\times 10^4\,\rm cm^{-3}} \right)  \\
     &\times&
     \left( \frac{M_{\rm core}}{1\,M_\odot} \right)^2
      \left( \frac{\sigma_{\rm cl}}{1\rm\,km\,s^{-1}} \right)^{-3}\,M_\odot\,{\rm yr^{-1}}
      \label{equ:bhl}
\end{eqnarray}
for the median clump density ($n_{\rm cl}$) of $4\times 10^4\,\rm cm^{-3}$ and the median velocity dispersion of the clumps ($\sigma_{\rm cl}$) of $1\rm\,km\,s^{-1}$ \citep{Morii23}. 
For cores with $M_{\rm core}>10\,M_\odot$, it will be on an order of 10$^{-5}$ $M_\odot$\,yr$^{-1}$. 
For an intermediate mass core ($\sim$10 $M_\odot$) to be a $M_{\rm core}$=30 $M_\odot$, the required duration time of prestellar phase is $\sim$1.5 Myr.  
The gray dashed line in Figure~\ref{fig:cmf_evolution} represents the case. 
The timescale needed for this model is comparable to the inertial inflow model proposed by \citet{Padoan20}, although it is much longer than the prestellar lifetime estimated in Section~\ref{subsec:lifetime}. 
The CMF became shallower as the accretion rate is proportional to mass squared, and it is much shallower than the observed protostellar CMF (red line). 

Then, to investigate which power-law index of mass can reproduce the observed CMFs, we assumed that the external mass infall rate follows a power-law dependence on mass ($\dot{M} = C M^{p}$), where C is a constant factor and estimated the best-fit parameter $p$ using non-linear least squares fitting with \texttt{scipy}'s \texttt{curve\_fit}.
The derived best-fit parameter is $p=1.16\pm0.02$ with $C=1.60\pm$0.01. 
Assuming a duration of external infall of 30 kyr, approximately free-fall time of ASHES cores, the mass infalll rate is described as 
\begin{equation}
    \dot{M}= 5.33\times 10^{-5} \left( \frac{M_{\rm core}}{1\,M_\odot} \right)^{1.16} \,M_\odot\,{\rm yr^{-1}}. 
    \label{equ:bestfit-mdot}
\end{equation} 
This model is plotted as green dashed line in Figure~\ref{fig:cmf_evolution}. 
Compared with BHL accretion, the constant factor, $C$, is much larger, and the smaller power-law index affects the core growth of initially low- to intermediate-mass cores, not only higher-mass cores.

We compare the estimated value with theoretical predictions. One is the clump-fed model derived by \citet{Tan14} as their Equation (16), which is 
\begin{eqnarray}
\nonumber
    \dot{M} &=& 6.26  \times 10^{-5} 
    \left(\frac{\epsilon_{\rm ff}}{0.1} \right) 
    \left(\frac{\epsilon_{\rm cl}}{0.5}\right)^{-1} \left(\frac{m_{*}}{50 \,M_\odot} \right)\\
    &\times&
    \left(\frac{\Sigma_{\rm cl}}{0.3 {\rm \,g\,cm^{-2}}}\right)^{3/4} 
    \left(\frac{M_{\rm cl}}{800 \,M_\odot}\right)^{-1/4} 
     \,M_\odot\,{\rm yr^{-1}},
\end{eqnarray} 
where $\epsilon_{\rm ff}$ is the star formation efficiency per free-fall time, $\epsilon_{\rm cl}$ is the final star formation efficiency from the clump, and $m_{*}$ is the final stellar mass. 
We inserted the median surface density and mass of the ASHES clump and assuming the $\epsilon_{\rm ff}$ and $\epsilon_{\rm cl}$ of 0.1 and 0.5, respectively, for normalization. 
The final stellar mass and the efficiencies are unclear but the constant factor and the power-law index is well similar to our best-fit, if the final stellar mass is somewhat proportional to the core mass. 

In the case of clump-fed model in a gas-dominated potential, where tidal-lobe accretion dominates, the mass accretion rate is expected to have a power-law index of 2/3 \citep{Bonnell01}, as discussed in \citet{Clark21}. 
Indeed, \citet{Maschberger14} confirmed the mass dependence by analyzing a hydrodynamical sink particle simulation following \citet{Bonnel11}. 
One recent observational study found $\dot{M} \propto M^{0.66}$ relation \citep{Wells24}, preferring the tidal-lobe accretion. 
The ASHES project targets the early stage of high-mass star formation where gas potential likely dominates, but the best-fit parameter implies a factor 2 larger index. 
If the external infall duration time depends on mass, where more massive stars form earlier, the index may become closer to 2/3 since we assumed a constant duration time. 

Equation~\ref{equ:bestfit-mdot} gives a much higher mass infall rate than BHL accretion (Equation~\ref{equ:bhl}).
Indeed, recent observations of core-scale infall have reported such high mass infall rates \citep[i.e., $\dot{M}>10^{-4}\,M_\odot\,{\rm yr}^{-1}$;][]{Contreras18, Sanhueza21, Redaelli22, Morii25, Sanhueza25}, which enable core growth in timescales comparable to freefall timescales.  
It indicates that the differential core growth may modify the Salpeter distribution produced by the initial fragmentation. 
In fact, the ALMA-IMF project has shed light on the evolution of CMF in massive clusters at different evolutionary stages \citep{Pouteau23}. 
Similarly, \cite{Coletta25-almagal} investigated the evolution of CMFs across three distinct evolutionary groups. Their findings indicate that CMFs in more evolved stages exhibit shallower tails and contain more massive cores, implying that core growth is a possible explanation. 

The Global and Local Infall in the ASHES Sample (GLASHES) survey reveal signs of a correlation between the core mass and the mass infall rate \citep{Morii25}. 
\citet{Wells24} also found a relation between core mass and flow rate along filamentary structures toward cores. 
These correlations will be further explored with a larger sample and can be interpreted as direct evidence of mass-dependent infall. 
This mechanism explains the change in the CMF slope at high masses with the passage of time, from the prestellar to the protostellar phase, as initially suggested by \citet{Sanhueza19}. 
It is also consistent with the idea that more massive cores have shorter lifetimes due to faster gas consumption (see Section~\ref{subsec:lifetime}). 

These findings suggest that core-to-star formation efficiency is not a straightforward process, challenging the assumption that initial core mass directly determines the final stellar mass. 
The evolution from a top-heavy CMF to Salpeter's IMF has not been discussed in this paper, primarily because our targets are in the similar evolutionary stages and the sample of protostellar cores is small.  
However, the ALMA-IMF project recently reported further evolution of the CMF from top-heavy to the Salpeter-like slope \citep[][and Cunningham+ in prep]{Armante24}.
Beyond core growth, processes such as further core fragmentation and multiple system formation may play a role \citep[e.g.,][]{Olguin22,Li24_natureas,Sanhueza25}. 
If the core-to-star formation efficiency depends on core mass, it also affect the evolution from CMF to IMF. 
\citet{Pouteau22} explored how the power-law slope of CMFs changes under various core-to-star formation efficiencies and fragmentation effects. 
Studying CMFs with a similar number of cores in evolutionary stages more advanced than the ASHES targets will allow for discussions similar to those of the ALMA-IMF project, which targeted more extreme and active environments. 

In summary, our study reveals a prestellar CMF with a power-law tail resembling Salpeter's IMF. 
Nevertheless, it is increasingly understood that the IMF is not a simple scaling of the CMF, and the process of mass accretion is considerably more complex. 

\section{Conclusion}
\label{sec:conclusion}
We have presented the CMF using 715 cores derived from the ASHES (ALMA Survey of 70 \textmu m dark High-mass clumps in Early Stages) project. 
We have obtained the following conclusions:

\begin{enumerate}
    \item The produced CMF has a power-law tail above $\sim$1 $M_\odot$. 
    The slope differs when the protostellar cores are included or excluded.
    When the protostellar cores are included, the CMF has a slope slightly shallower than -2.35. 
    In contrast, the CMF from only the prestellar cores has a steeper slope due to the lack of higher-mass prestellar cores.  
    
    \item The estimated power-law slopes of the CMF of prestellar core candidates above the completeness limit of 1.2 $M_\odot$ were -2.65 to -2.38, depending on the CMF representation and fitting method. 
    The CMF of the gravitationally bound prestellar core has a slightly shallower slope of -2.53 to -1.83. 
    Both are indistinguishable from Salpeter's slope of -2.35. 
    
    \item The power-law index of the prestellar CMFs is similar to those derived in low-mass star-forming regions and also in more evolved, massive protoclusters. This implies a universal core formation mechanism to produce the mass distribution between low- and high-mass star-forming regions. 

    \item We estimated the lifetime of prestellar cores in IRDCs based on the fraction of prestellar cores against the protostellar cores. This lifetime decreases as the core mass increases and approaches the freefall time at $M_{\rm core} \sim$10 $M_\odot$. Considering the core growth, even in the high-mass range, the lifetime of prestellar cores becomes a few times the cores' freefall time, similar to the statistical lifetime estimated in low-mass star forming regions. 

     \item The top-heavy, shallower CMF, including protostellar cores, observed in the ASHES sample is similar to the one seen in massive protoclusters. The core growth in mass with a higher mass infall rate and a smaller exponential index than the Bondi-Hoyle-Lyttleton accretion rate has the potential to produce such a shallower slope and increase in the core mass from the prestellar to protostellar stage. 
\end{enumerate}

\begin{acknowledgments}
We are grateful to the anonymous referee for their constructive suggestions, which helped improve the quality of this paper. 
K.M was financially supported by Grants-in-Aid for the Japan Society for the Promotion of Science (JSPS) Fellows (KAKENHI No. 22J21529). 
K.M was supported by the ALMA Japan Research Grant of NAOJ ALMA Project, NAOJ-ALMA-369.
PS was partially supported by a Grant-in-Aid for Scientific Research (KAKENHI Number JP23H01221) of the Japan Society for the Promotion of Science (JSPS). 
GS acknowledges the project PRIN MUR 2022 FOSSILS (“Chemical origins: linking the fossil composition of the Solar System with the chemistry of protoplanetary disks”, Prot. 2022JC2Y93), the project ASI-Astrobiologia 2023 MIGLIORA (“Modeling Chemical Complexity”, F83C23000800005), the INAF-GO 2023 fundings PROTOSKA (“Exploiting ALMA data to study planet forming disks: preparing the advent of SKA”, C13C23000770005) and the INAF Minigrant 2023 TRIESTE (“TRacing the chemIcal hEritage of our originS: from proTostars to planEts”; PI: G. Sabatini). 
X.L. acknowledges support from the Strategic Priority Research Program of the Chinese Academy of Sciences (CAS) Grant No.\ XDB0800300, the National Key R\&D Program of China (No.\ 2022YFA1603101), State Key Laboratory of Radio Astronomy and Technology (CAS), the National Natural Science Foundation of China (NSFC) through grant Nos.\ 12273090 and 12322305, the Natural Science Foundation of Shanghai (No.\ 23ZR1482100), and the CAS ``Light of West China'' Program No.\ xbzg-zdsys-202212.
Data analysis was in part carried out on the Multi-wavelength Data Analysis System operated by the Astronomy Data Center (ADC), National Astronomical Observatory of Japan. 
This paper uses the following ALMA data: 
ADS/JAO.ALMA\#2015.1.01539.S, ADS/JAO.ALMA\#2017.1.00716.S and ADS/JAO.ALMA\#2018.1.00192.S (PI: P. Sanhueza). 
ALMA is a partnership of ESO (representing its member states), NSF (USA), and NINS (Japan), together with NRC (Canada), $MOST$ and ASIAA (Taiwan), and KASI (Republic of Korea), in cooperation with the Republic of Chile. The Joint ALMA Observatory is operated by ESO, AUI/NRAO, and NAOJ.
Data analysis was in part carried out on the open-use data analysis computer system at the Astronomy Data Center (ADC) of the National Astronomical Observatory of Japan. 
\end{acknowledgments}

\facility {ALMA} 
\software{Astropy \citep{Astropy_13, Astropy18, Astropy22}, astrodendro \citep{Rosolowsky08}, CASA \citep[][]{CASA22}, getsf \citep{Menshchikov13, Menshchikov21}, Matplotlib \citep{Hunter_matplotlib}, Numpy \citep{Harris20_numpy}, powerlaw \citep{Alstott14}, Scipy \citep{Virtanen_scipy},  seaborn \citep{Waskom2021_seaborn}.} 

\appendix

\section{Core Mass Distribution by Evolutionary Stage}
\label{sec:Appendix_Ncore} 
\begin{figure}
    \centering
    \includegraphics[width=0.8\linewidth]{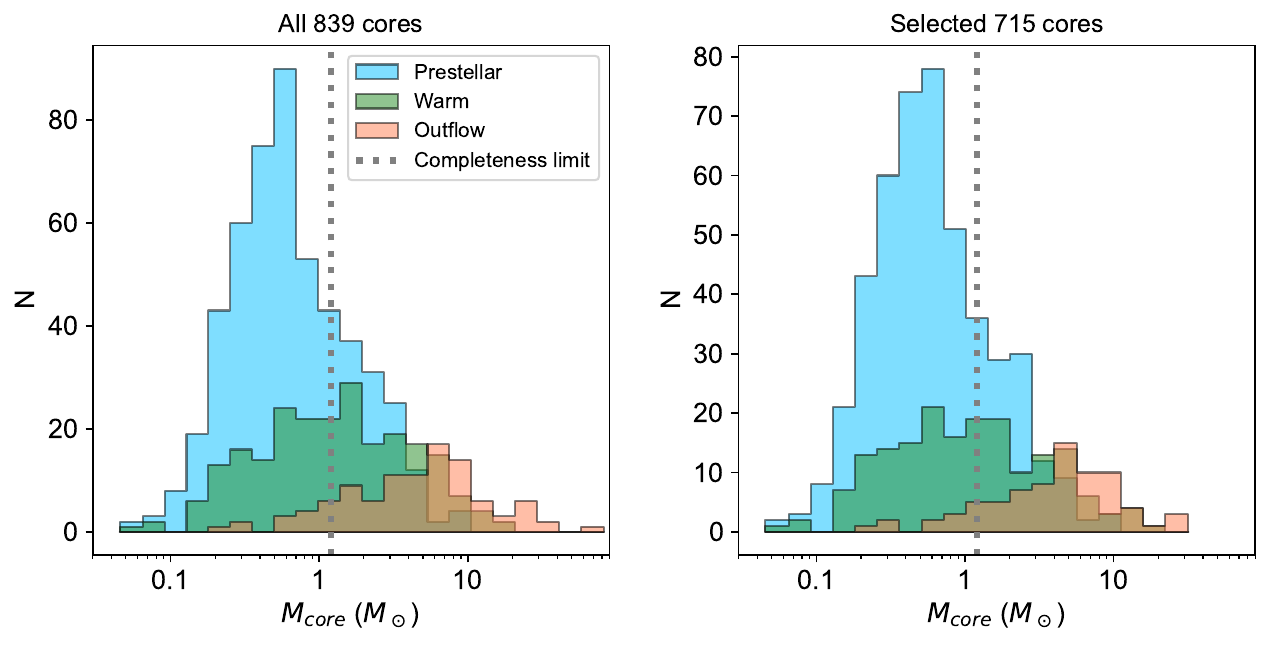}
    \caption{Number of cores in each mass bin, separated by evolutionary stage. Light blue, green, and orange represent prestellar core candidates, warm cores, and outflow cores, respectively. The left panel shows all ASHES cores, while the right panel shows the selected 715 cores (see Section~\ref{sec:completeness}). The vertical dotted line indicates the mass completeness limit.}
    \label{fig:Ncore}
\end{figure}

Figure~\ref{fig:Ncore} shows the number of cores in each mass bin for each evolutionary stage.
In contrast to Figure~\ref{fig:cmf-all}, the histograms are not stacked and the y-axis represents the pure number count, allowing for a clearer view of the population in each evolutionary stage as a function of core mass.
The left and right panels show the distributions for all 839 cores and the selected 715 cores, respectively.
This figure clearly demonstrates that we have detected outflow and warm gas tracers even from low-mass cores well below the completeness limit of 1.2~$M_\odot$ (gray line).

\section{Core Identification: astrodendro VS getsf}
\label{sec:Appendix_coremethod}
Cores are not an obvious structure.
We define compact, dense structures that eventually collapse to form a single star or a multiple stellar system as cores.
However, they are usually embedded in clumps or clouds, and sometimes they do not have clear boundaries from the surroundings. 
Therefore, there is always uncertainty when setting parameters and using algorithms for identification.
Additionally, when identifying cores from observational images, we must keep in mind projection effects.

There are some identification algorithms or techniques such as \texttt{GaussClump} \citep{StutzkiGuesten90_gaussclump}, \texttt{ClumpFind} \citep{Williams94}, \texttt{CuTEx} \citep[Curvature Thresholding Extractor;][]{Molinari11}, \texttt{CSAR} \citep[Cardiff Sourcefinding AlgoRithm;][]{kirk13_csar}, \texttt{getsources} or \texttt{getsf} \citep{Menshchikov13, Menshchikov21}, \texttt{FellWalker} \citep{Berry15}, \texttt{GExt2D} (Bontemps et al., in prep.), as well as \texttt{astrodendro} \citep{Rosolowsky08}. 
Each core (or clump) identification method has pros and cons. 
Some examples of discussion are given in \citet{Joncour20_coreexcraction, LiChong20_identification, Menshchikov21, Cheng24}, and \citet{Pineda23}.
\citet{LiChong20_identification} demonstrated that the \texttt{astrodendro} has higher detection completeness and better performance in extracting parameters (e.g., size, mass, and flux) than other algorithms such as \texttt{ClumpFind} and \texttt{GaussClumps}.

As the continuum images of the ASHES project show, the clumps have pc-scale internal structures and compact sources embedded. 
In such a hierarchical structure, we tried to extract only the compact components and applied \texttt{astrodendro}. 
However, here as an appendix, we show how the different algorithms give different outputs by comparing \texttt{astrodendro} and \texttt{getsf}. 

\begin{figure}
    \centering
    \includegraphics[width=\linewidth]{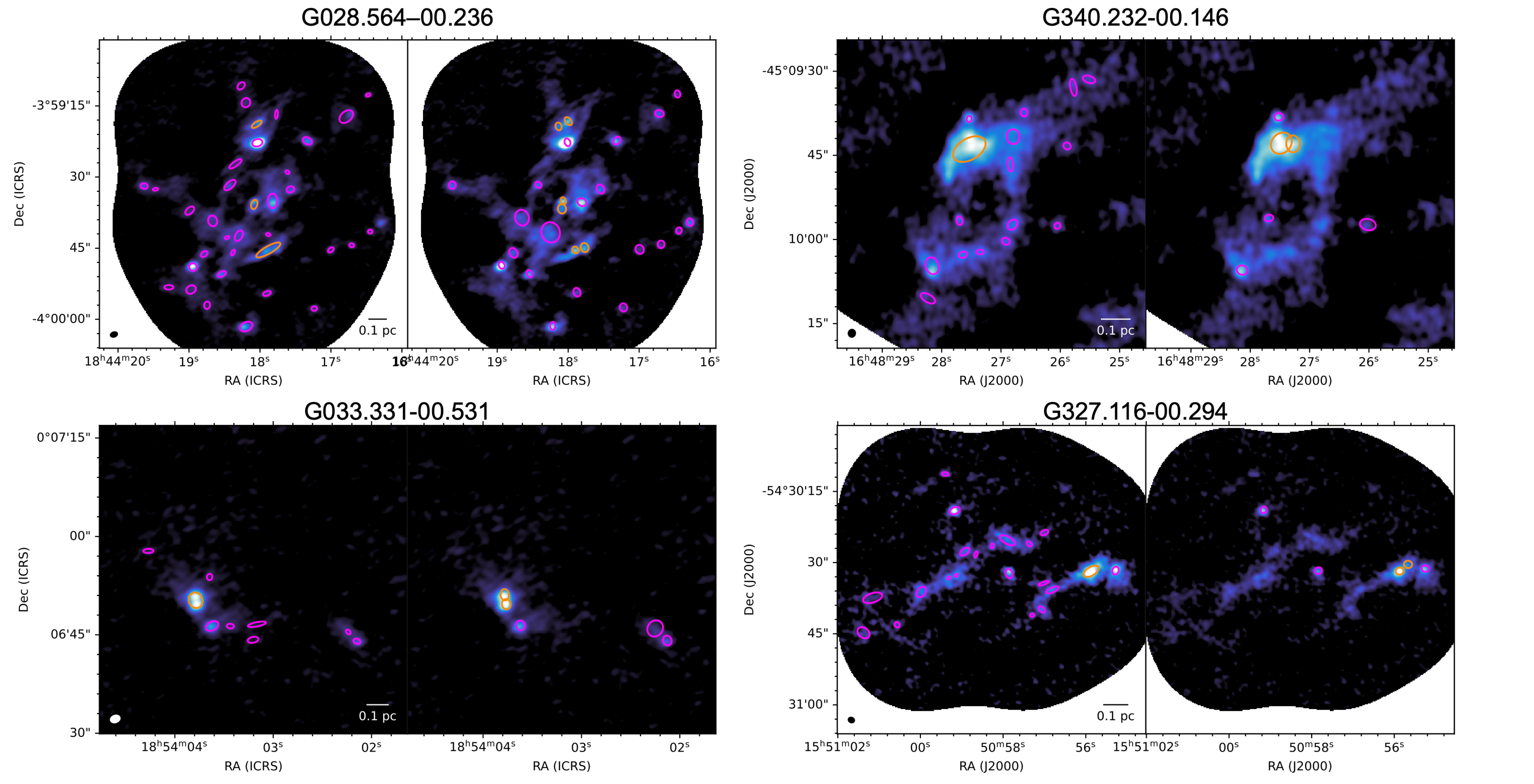}
    \caption{Comparison of the identified structures between \texttt{astrodendro} and \texttt{getsf}. These are examples of G028.564-00.236, G340.232-00.146, G033.331-00.531, and G327.116-00.294, from top left to bottom right. In each region, contours represent the structures identified by \texttt{astrodendro} and \texttt{getsf} in the left and right panels, respectively. The \texttt{astrodendro} cores where two \texttt{getsf} sources were identified are highlighted in orange. The background is the 1.3 mm continuum emission from \citet{Morii23}. }
    \label{fig:image_dendro_getsf}
\end{figure}

As we did with \texttt{astrodendro}, we ran \texttt{getsf} for 39 continuum images.
In this case, we only needed to set the maximum size of the sources, which did not significantly affect the output.
We set it to 10 arcsec, which is almost ten times larger than the beam size.
As a result, 331 cores were identified from 39 clumps.
Of those, 305 overlap with the cores identified by \texttt{astrodendro} according to their positions. 
The other 26 structures are only identified by \texttt{getsf}, though most are at the edge of the field of view or very small. 
There is no parameter regarding the minimum size of a structure for \texttt{getsf} identification, so it found further fragmentation in twelve cores identified by the dendrogram. 
Figure~\ref{fig:image_dendro_getsf} shows examples of differences in identified structures. 
Magenta contours indicate the structures identified by each method in the left (\texttt{astrodendro}) and right (\texttt{getsf}) panels.
Regions where a single \texttt{astrodendro} source corresponds to two distinct \texttt{getsf} sources are highlighted in orange. 

\begin{figure}
    \centering
    \includegraphics[width=0.8\linewidth]{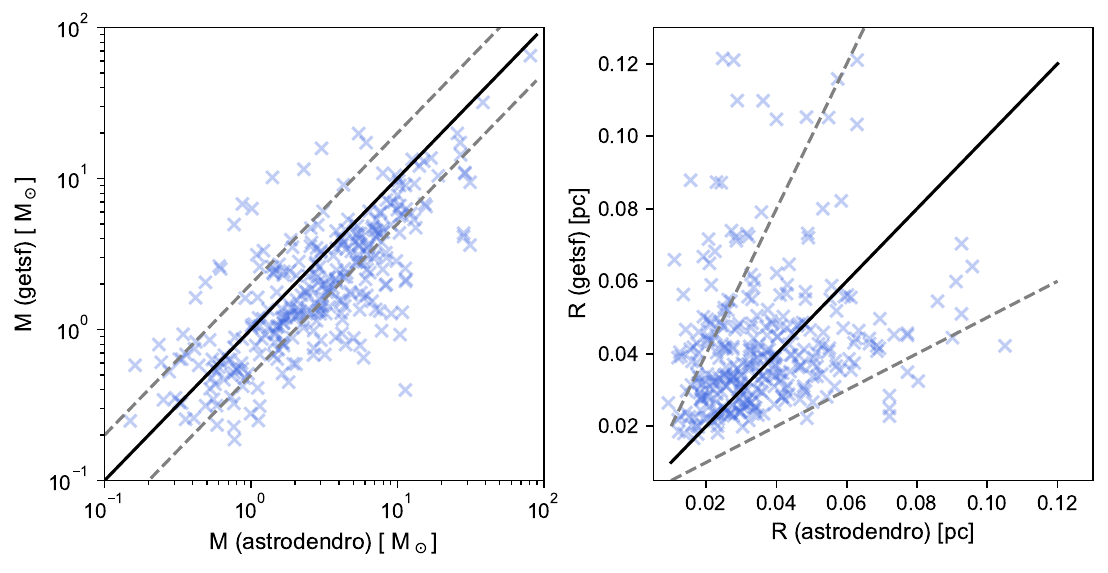}
    \caption{Difference of Mass and Size between \texttt{astrodendro} (x-axis) and \texttt{getsf} (y-axis). Black lines correspond to 1:1 correspondence, and gray lines represent factor 2 differences, or y= (2 or 0.5) $\times$ x. }
    \label{fig:MR_dendro_getsf}
\end{figure}

To quantitatively compare the physical properties, we plot the mass and the radius (geometrical mean of the major and minor FWHM) estimated by \texttt{getsf} and \texttt{astrodendro} in Figure~\ref{fig:MR_dendro_getsf}. 
Only cores that overlap with both samples (305) are plotted here.
The plot clearly shows that the size of cores measured from \texttt{astrodendro} are generally larger, and the flux and mass are also larger, though the differences are mostly within a factor of two. 

\begin{figure}
    \centering
    \includegraphics[width=0.6\linewidth]{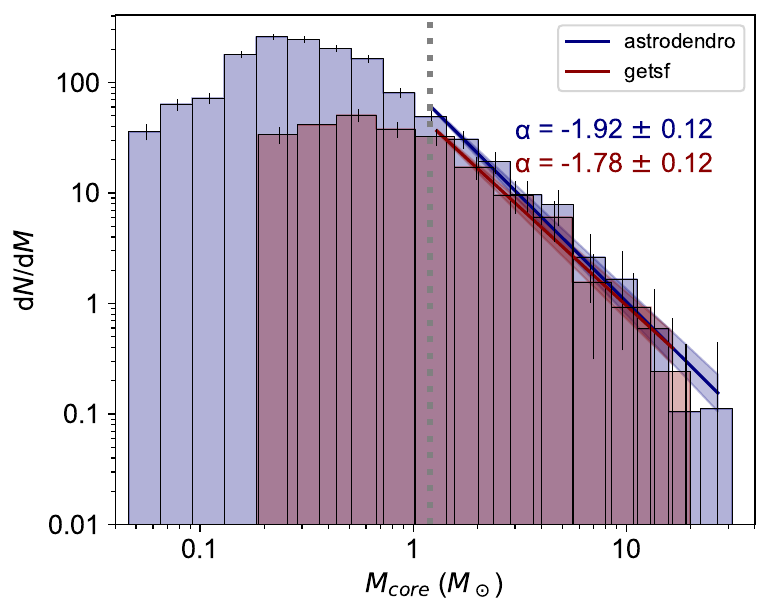}
    \caption{Core mass function of cores identified by \texttt{astrodendro} (navy) and \texttt{getsf} (dark red). The power-law fitting was applied to the sample with $M > 1.2 M_\odot$ as following the analysis in Section~\ref{sec:completeness}.}
    \label{fig:CMF_dendro_getsf}
\end{figure}

Then, to see how significantly the mass function differs, we made a CMF for both samples. 
Figure~\ref{fig:CMF_dendro_getsf} shows the difference by plotting the CMFs with the same bins.  
The navy and dark red colors highlight the different samples of \texttt{astrodendro} and \texttt{getsf}, respectively. 
As in the main text, we excluded five clumps here. The navy plot is the same as the middle panel of Figure ~{fig:cmf-all}. 
The power-law fit was applied to the CMFs with $M > 1.2 M_\odot$, and the best-fit parameters are $\alpha = -1.92\pm0.12$ and $-1.78\pm0.12$ for \texttt{astrodendro} and \texttt{getsf}, respectively. 
This indicates that individual core masses are not always similar, but the shallower slope than -2.35 holds and does not significantly change our conclusion in the analysis of CMFs.

\bibliography{references}
\bibliographystyle{aasjournalv7}

\end{document}